\documentclass[letterpaper]{article} 
\usepackage{aaai24}  
\usepackage{times}  
\usepackage{helvet}  
\usepackage{courier}  
\usepackage[hyphens]{url}  
\usepackage{graphicx} 
\usepackage{natbib}  
\usepackage{caption} 
\usepackage{algorithm}
\usepackage{algorithmic}

\usepackage{amssymb}
\usepackage{amsmath}
\usepackage{newfloat}
\usepackage{listings}
\DeclareCaptionStyle{ruled}{labelfont=normalfont,labelsep=colon,strut=off} 
\floatstyle{ruled}
\newfloat{listing}{tb}{lst}{}
\floatname{listing}{Listing}
\title{MLP-AMDC: An MLP Architecture for Adaptive-Mask-based Dual-Camera snapshot hyperspectral imaging}
\author{
    Zeyu Cai, Can Zhang, Xunhao Chen, Shanghuan Liu, Chengqian Jin, Feipeng Da
}
\affiliations{
    Southeast University, Nanjing, china;
    Chinese Academy of Agricultural Sciences, Nanjing, china.
}

\usepackage{bibentry}

\begin{document}

\maketitle
\begin{abstract}
Coded Aperture Snapshot Spectral Imaging (CASSI) system has great advantages over traditional methods in dynamically acquiring Hyper-Spectral Image (HSI), but there are the following problems. 1) Traditional mask relies on random patterns or analytical design, both of which limit the performance improvement of CASSI. 2) Existing high-quality reconstruction algorithms are slow in reconstruction and can only reconstruct scene information offline. To address the above two problems, this paper designs the AMDC-CASSI system, introducing RGB camera with CASSI based on Adaptive-Mask as multimodal input to improve the reconstruction quality. The existing SOTA reconstruction schemes are based on transformer, but the operation of self-attention pulls down the operation efficiency of the network. In order to improve the inference speed of the reconstruction network, this paper proposes An MLP Architecture for Adaptive-Mask-based Dual-Camera (MLP-AMDC) to replace the transformer structure of the network. Numerous experiments have shown that MLP performs no less well than transformer-based structures for HSI reconstruction, while MLP greatly improves the network inference speed and has less number of parameters and operations, our method has a 8 db improvement over SOTA and at least a 5-fold improvement in reconstruction speed. (https://github.com/caizeyu1992/MLP-AMDC.)
\end{abstract}

\section{Introduction}
\begin{figure}[t]
  \centering
   \includegraphics[width=1\linewidth]{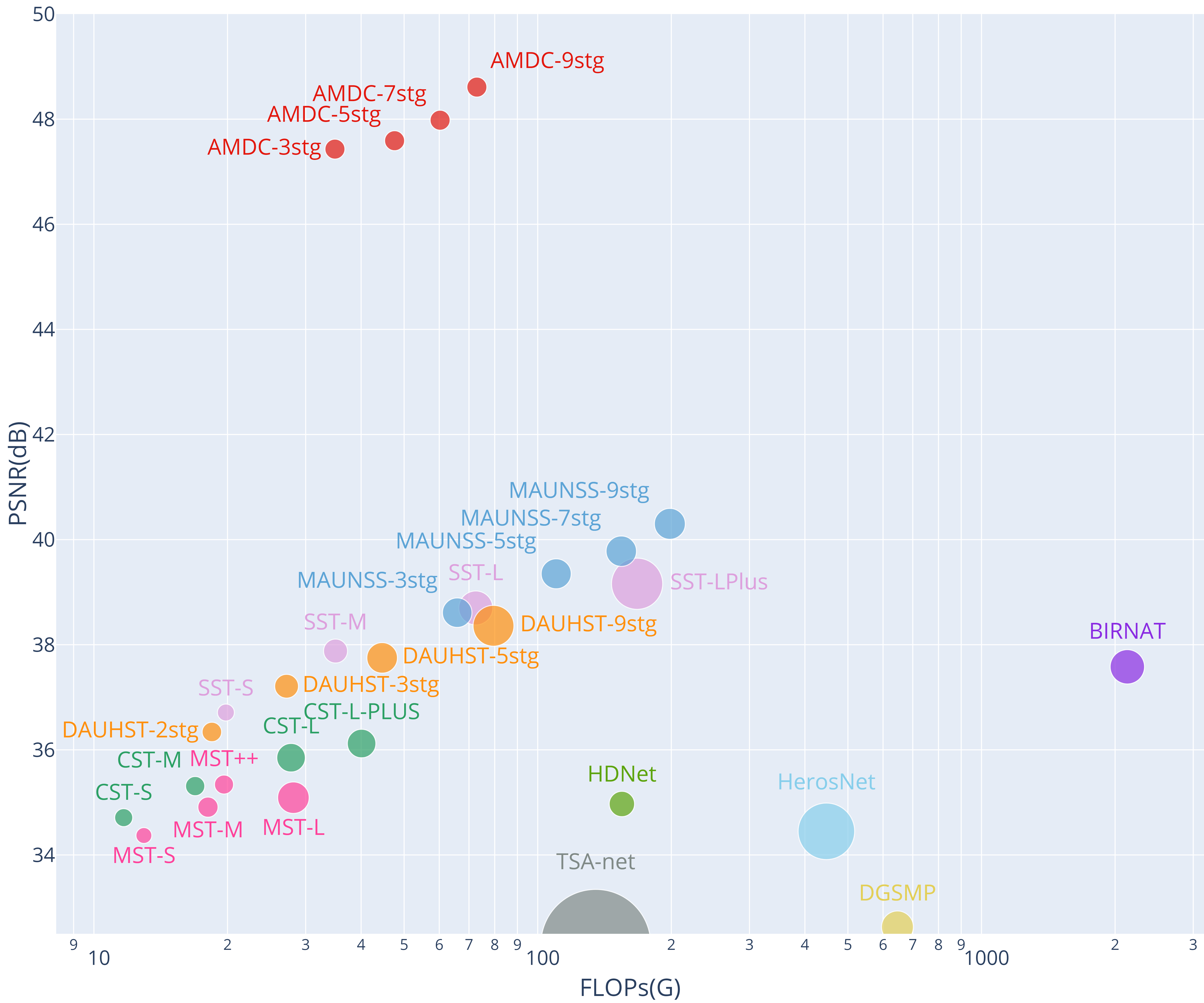}
   \caption{PSNR-parameters-FLOPs comparisons of our MLP-AMDC and SOTA HSI reconstruction methods.The vertical axis is PSNR (dB), the horizontal axis is FLOPs (computational cost), and the circle radius is parameters (memory cost).}
   \label{fig:1}
\end{figure}

Hyper-Spectral Image (HSI) has important applications in astrophysics, remote sensing, precision agriculture, medical biology, material identification, photogrammetry and other fields \cite{schechner2021guest, guo2022deep}. Conventional hyper-spectral imaging uses line sweep, surface sweep, or rotating filters to capture HSI 3D cube. the introduction of mechanical components has resulted in a less reliable system and is not capable of capturing dynamic scenes. CASSI based on compressed sensing has theoretically demonstrated the advantages of CASSI: 1) the ability to capture HSI of dynamic scenes\cite{cao2016computational}, 2) lower storage cost, and 3) more reliable lifetime without mechanical components.

However, current CASSI systems and reconstruction algorithms have two problems. 1) The MASK in CASSI either relies on random patterns or analytical design, and although existing algorithms achieve good results in these Masks, the fixed coding is not adaptive to changes in the data and more detailed textures are lost. In addition, the joint design of coding elemnets (CEs) and computational decoders is a trend in computational optical imaging (COI). Is it possible to design a multimodal adaptive coding in spectral compressed imaging to compensate for the lack of fixed coding? 2) Although the CASSI reconstruction algorithm achieves good reconstruction results, the reconstruction speed cannot meet the requirements of online reconstruction, and the existing algorithms are still offline reconstruction of HSI. how to improve the reconstruction speed without reducing the reconstruction accuracy?

To solve above problems:
To further improve the reconstruction quality, in this paper, we firstly introduce Adaptive-Mask to adjust the Mask by learning the distribution of data to improve the capture capability of CASSI. Secondly, in order to reconstruct the lost detailed texture information, we introduce RGB cameras and design a dual-stream network of RGB images and CASSI images to complement each other with information from different modalities.
To improve the reconstruction speed, we propose an multi-layer perceptron (MLP) Architecture for the reconstruction of spectral images, which does not use self-attention. Instead, the new architecture is based entirely on MLPs that are repeatedly applied across either spatial channels or spectral channels. To evaluate the performance of different algorithms in terms of reconstruction speed, we introduce frame rate as an evaluation metric on the dataset.

The mian contributions of this work can be summarized as follows:

\begin{itemize}
\item We propose a dual-camera CASSI system based on Adaptive-Mask, a multimodal CASSI system with joint optimization of CEs and decoding networks.
\item We designed an MLP Architecture for hyperspectral reconstruction, and designed spectral-MLP and Swin-spatial-MLP to obtain the global correlation between CASSI measurements and RGB images, respectively, and embed them into AMDC CASSI.
\item We demonstrate on several datasets that MLP, a simple structure, has no less performance than CNN, Transformer, and has faster inference, including a larger and more recent dataset (ARAD\_1K). Numerous experiments have shown that our method has faster reconstruction results and reconstruction speed, and has less number of parameters and operations.
\end{itemize}

\section{Related Work}
\subsection{Related works of CASSI system coding}

\begin{figure}[t]
  \centering
   \includegraphics[width=0.9\linewidth]{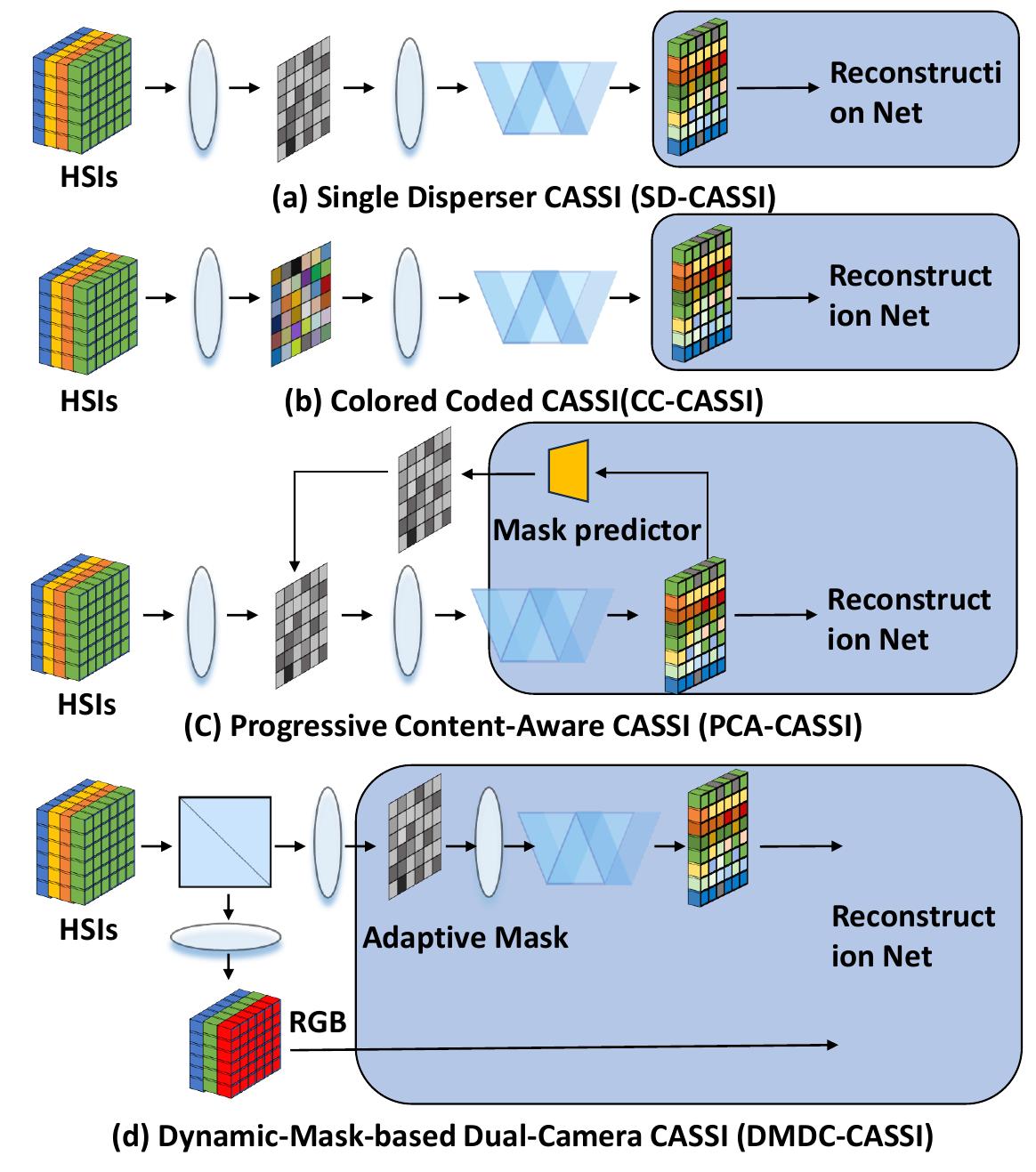}
   \caption{Detailed architecture of our MLP-AMDC.}
   \label{fig:1}
\end{figure}

The coding development of CASSI systems is shown in Fig. 2. A typical Single-Dispersive CASSI (SD-CASSI) design is shown in Fig. 2(a) \cite{wagadarikar2008single}, SD-CASSI utilizes a manual a priori grayscale Mask to encode the HSI, which is subsequently disperse by a dispersive device and eventually measured by the sensor. Then Dual-Dispersive CASSI (DD-CASSI) with spectral-spatial encoding \cite{gehm2007single} and Colored Coded CASSI (CC-CASSI, Fig. 2(b)) \cite{arguello2014colored} with multi-channel spectral encoding have been developed. Previous studies have focused on how to improve the encoding efficiency with a fixed Mask, without much consideration of the adaptation of the Mask to the data. In order to improve the effect of fixed MASK on CASSI, a multi-MASK CASSI system (MS-CASSI) at the expense of CASSI multiple imaging has been proposed \cite{kittle2010multiframe}. Besides, Some existing works on traditional CS have explored the possibility of joint mask optimization and image reconstruction. For instance, Zhang et al. proposed a constrained optimization-inspired network (HerosNet) for adaptive sampling and recovery \cite{zhang2022herosnet}. In the spectral SCI, Zhang et al. designed an end-to-end learnable auto-encoder to optimize the illumination pattern and compress the HSIs \cite{zhang2023progressive}, as shown in Fig. 2(c). However, there are fewer studies on the methods of adaptive Mask for CASSI enhancement, and the multimodal network composed of adaptive-Mask based CASSI with RGB camera inspired by DCCHI \cite{wang2016adaptive} has not been discussed yet. The study of joint optimization of adaptive MASK and multimodal networks is still challenging and worth exploring.

\subsection{HSI reconstruction algorithms}
CASSI projects HSI from 3D cube to 2D camera CCD sensor, the reconstruction problem is an ill-posed problem. HSI reconstruction algorithms can be classified into traditional methods, plug-and-play methods, End-to-End (E2E) methods, deep unfolding frameworks, and optical path reversible-based frameworks.

\begin{figure*}[t]
  \centering
   \includegraphics[width=0.82\linewidth]{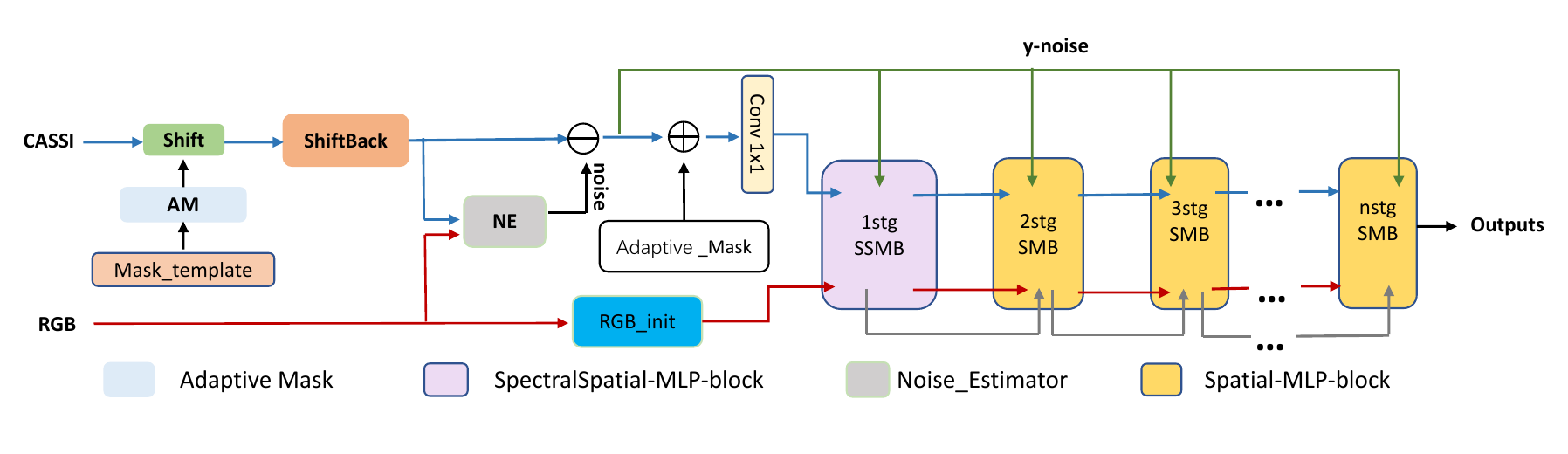}
   \caption{Detailed architecture of our MLP-AMDC.}
   \label{fig:2}
\end{figure*}

\begin{figure*}[t]
  \centering
   \includegraphics[width=0.85\linewidth]{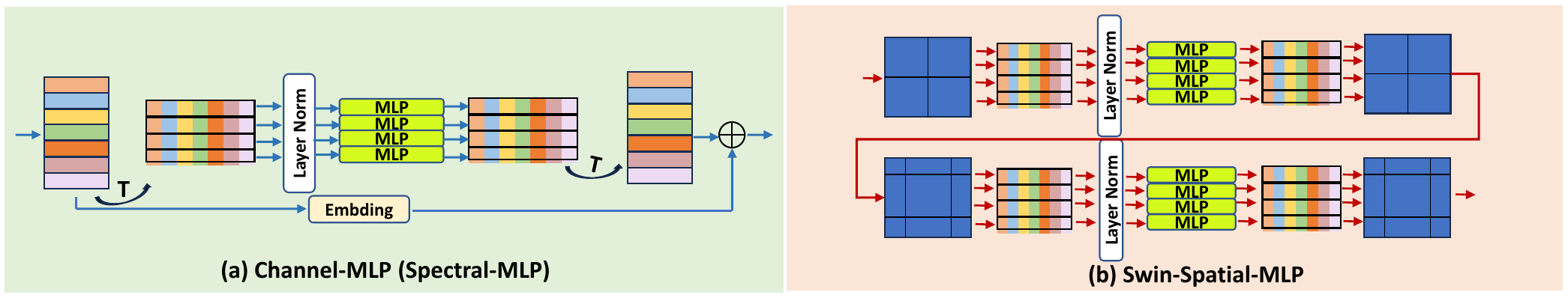}
   \caption{Detailed architecture of our MLP-AMDC.}
   \label{fig:3}
\end{figure*}

Traditional methods are based on optimization theory and are solved by iteratively minimizing the objective function \cite{liu2018rank, wang2021tensor, bioucas2007new}. For example, Yuan et al. used generalized alternating projection (GAP) to solve the total variation (TV) minimization problem based on the CS theory \cite{yuan2016generalized}. Plug-and-play methods improve the reconstruction quality by introducing deep learning-based denoising modules into traditional methods. E2E methods are classified into CNN-based, RNN-based, and Transformer-Based methods, e.g., Zhang et al. trained a convolutional neural network (CNN)-based coded HSI reconstruction method by coded image internal learning after external learning by public HSI datasets \cite{zhang2019hyperspectral}. Yang et al. introduced a symmetric residual module and a nonlocal spatial–spectral attention module into a deep learning network to capture nonlocal spatial and spectral correlations of HSI \cite{yang2021hyperspectral}.  Miao et al. introduced self-attention to model nonlocal features \cite{miao2019net}. Cai et al. proposed a spectral global feature extraction module of MSA-attention to obtain good reconstruction results with low parameters \cite{cai2022mask}. The depth unfolding framework replaces the convex optimization formulation by the depth module and obtains interpretable results by iterative solving. The reversible optical path-based framework reprojects each reconstruction result to the 2D measurement space, constructs new residuals with the measured values, and then minimizes the residuals by iterations to obtain results comparable to the depth unfolding framework, while having fewer parameters and operations \cite{cai2022degradation}.

\section{Methods}
\subsection{Model of AMDC-CASSI System}

The concise AMDC-CASSI schematic is shown in Fig. 2(d). In terms of math, think about a 3D HSIs cube, denoted by $X\in \mathbb{R} ^{n_{x}\times n_{y}\times n_{\lambda } }$, where $n_{x}$, $n_{y}$, $n_{\lambda }$ represent the HSIs’s height, width, and number of wavelengths. First, the scene goes through a beamsplitter. By default, the energy of the light is split evenly, and the RGB measurement, g(x,y), can be written as:

\begin{equation}
\setlength{\abovedisplayskip}{0pt} 
 g\left ( x,y \right ) =\frac{1}{2} \int_{\Lambda }\omega \left ( \lambda  \right ) X\left ( x,y,\lambda  \right )d\lambda +  N 
  \label{eq:CASSI0}
\end{equation}
where$x\in n_{x},y \in n_{y},\lambda \in n_{\lambda}$, $X\left (x,y,\lambda \right )$ is the 3D hyperspectral data cube. $\Lambda$ is the spectral response range of the RGB detector. $\omega \left ( \lambda  \right )$ is the spectral response function of the detector. $N$ is the noise on the RGB detector.

The RGB measurements are passed through an adaptive Mask network that learns the spatial features of the scene and produces an adaptive Mask, which can be expressed as:

\begin{equation}
\setlength{\abovedisplayskip}{0pt} 
 M\left ( x,y \right ) =AdaptiveMaskNet\left ( g(x,y) \right ) 
  \label{eq:CASSI0}
\end{equation}

Where $M$ denotes the Mask of CASSI. The other light path of the spectroscopic prism, which is then modulated by the adaptive mask and a dispersive prism, is finally captured by the CASSI detector. We can express it as:
\begin{equation}
\setlength{\abovedisplayskip}{0pt} 
 X^{\prime } (x,y,: ) = \frac{1}{2}M(x,y)\odot X(x,y,:)
  \label{eq:CASSI0}
\end{equation}
Where $X^{\prime } \in \mathbb{R}^{n_{x}\times n_{y} \times n_{\lambda }}  $ denotes the modulated spectral data cub.

\begin{equation}
\setlength{\abovedisplayskip}{0pt} 
 X^{\prime\prime } (x,y,: ) = X^{\prime }(x,y+d_{\lambda},:)
  \label{eq:CASSI0}
\end{equation}

Finally, the captured 2D compressed measurement $Y\in \mathbb{R} ^{n_{x}\times \left ( n_{y}+d\left ( n_{\lambda } -1 \right )   \right )  }$ can be obtained by:

\begin{equation}
\setlength{\abovedisplayskip}{0pt} 
Y =\frac{1}{2}\sum_{\lambda=1}^{n_{\lambda } } M(x,y+d_{\lambda})\odot X(x,y+d_{\lambda},:) + G
  \label{eq:CASSI2}
\end{equation}
where $G\in \mathbb{R} ^{n_{x}\times \left ( n_{y}+d\left ( n_{\lambda }-1  \right ) \right ) }$ is the random noise generated by the photon sensing detector during the measurement.

Combining Eq. (1) and Eq. (5) and rewriting them as linear transformations, the imaging model of dual-camera in the matrix form can be written as:

\begin{equation}
\setlength{\abovedisplayskip}{0pt} 
\begin{cases}
 Y_{r} = \Phi _{r} X +N_{r} 
\\
Y_{c} = \Phi _{c} X +N_{c}
\end{cases}
  \label{eq:CASSI0}
\end{equation}
Where $Y_{r}$ is the measurement of RGB detector. $N_{r}$ is the noise of RGB detector. $\Phi_{r}$ is the sensing matrix  of RGB detector. $Y_{c}$ is the measurement of CASSI detector. $N_{c}$ is the noise of CASSI detector. $\Phi_{c}$ is the sensing matrix of CASSI detector, which is generally consider it as the shifted mask.

\subsection{Overall architecture of MLP-AMDC}

The observational model in Eq. (6) classifies HSI reconstruction from AMDC-CASSI and panchromatic images as an optimization problem. An observation model-derived regularization model is recommended:

\begin{equation}
\setlength{\abovedisplayskip}{0pt} 
\arg\min \left \| Y_{c}-N_{c} -\Phi _{c}X \right \|_{2}^{2} +  \left \| Y_{r}-N_{r} -\Phi _{r}X \right \|_{2}^{2}
  \label{eq:CASSI0}
\end{equation}

Equation (6) says each detector has noise. We assume that the sensor noise has the same distribution, which can be inferred from both detectors.

\begin{equation}
\setlength{\abovedisplayskip}{0pt} 
noise =\frac{\Phi_{c}^{-1}Y_{c} -\Phi_{r}^{-1}Y_{r}}{\Phi_{c}^{-1}-\Phi_{r}^{-1}} =Denoise(Y_{c},Y_{r})
  \label{eq:CASSI0}
\end{equation}

During reconstruction, based on the optical path's invertibility, we reproject each reconstruction's results back into measurement space, inspired by SST-net \cite{cai2023sst}. The unique solution of the forward process and the actual measurements can create residuals for the network to learn the detailed features of the spectral cube, as described by the following equation:

\begin{equation}
\setlength{\abovedisplayskip}{0pt} 
x_{n+1} = f(Y_{c}-N-\Phi _{c}x_{n} ,Y_{r} )+x_{n} 
  \label{eq:CASSI0}
\end{equation}

where $x_n$ is the result of the reconstruction after the n-stage iteration. $f$ is an E2E reconstruction module that shares weighted parameters in different phases, except for the initialization phase of the reconstruction.

\subsection{Adaptive mask network}
As illustrated in Fig. 3, the adaptive mask network learns the spatial distribution of the scene from the dataset during training and predicts if each pixel represents redundant information. we designed the CNN-based network using the manual mask as a template. This adaptive mask network will update weights synchronously with the reconstruction network during training, then freeze and update only the reconstruction network weights, save the Mask, and delete this module during inference. Adaptive mask network:

\begin{equation}
\setlength{\abovedisplayskip}{0pt} 
M(x,y) =\mathcal{S} (\mathcal{C} (\mathcal{U}(y_{rgb}(x,y) + \mathcal{MT}(x,y))))
  \label{eq:CASSI0}
\end{equation}
where $\mathcal{S}$ is Sigmoid function, $\mathcal{C}$ is a Convolution layer, $\mathcal{U}$ is a u-net, $\mathcal{MT}$ is the Mask Template.

\subsection{Multimodal reconstruction network}
The function of the multimodal reconstruction network is to reconstruct the spectral 3D cube from the RGB and CASSI measurements. As shown in Fig. 3, the reconstruction network uses a dual-stream architecture and constructs a residual learning network of Eq. (9) based on the reversible nature of the optical path, and reconstitutes the input of n+1 stage using the results of the n-stage to improve the reconstruction quality. The key modules of the Subnet are Noise-Estimator module (NE), RGB-init module, Swin-Spatial-MLP block and SpectralSpatial MLP block. First, the network predicts the noise distribution of the detector by the NE module and corrects the CASSI data. Subsequently, the job of the RGB initialization module is to transfer RGB to a higher dimensional space, synchronize the number of RGB channels with the number of CASSI channels, and assist in learning the long-range correlations of RGB branches in each channel. Finally, the dual-stream data are iterated to learn the entire correlation of the spectrum and the full correlation of the space.

\noindent\textbf{Noise-Estimator module (NE).} 
We presume that the distribution function of the noise on the RGB detector and the CASSI detector is the same. The NE module transfers the RGB and CASSI branches to a high-dimensional space, learns the noise distribution, downscales it to the original space, and then estimates the noise using the imbalance data between the two branches. The following is a description of this procedure:

\begin{equation}
\setlength{\abovedisplayskip}{0pt} 
N= \mathcal{O}(\mathcal{D}(\mathcal{C}(Up(x_{c})))) -\mathcal{O}(\mathcal{D}(\mathcal{C}(Up(x_{r})))) +\varepsilon
  \label{eq:CASSI0}
\end{equation}
where $N$ is the estimated noise, $\mathcal{O}$ is a Softmax layer, $\mathcal{D}$ is a downsampling layer, $\mathcal{C}$ is a Convolution layer,  Up is a upsampling layer.

\noindent\textbf{RGB-init module.} RGB initialization module that scales raw RGB measurements to 64 dimensions before downscaling them to original channels, allowing RGB measurements to be aligned with CASSI channel counts.

\noindent\textbf{SpectralSpatial MLP block.} Because C<<W=H, Channel-MLP has the lowest reconstruction cost and is suited for CASSI branch and RGB branch initiation. Swin-Spatial-MLP learns the global similarity of the two branches in space using spectral characteristics.

\noindent\textbf{Spatial MLP block.} The residuals of the measurement and the n-stage reconstruction results are constructed, the Swin-Spatial-MLP is utilized to learn the global correlation between the residuals and the RGB images in space.

\subsection{Loss function}
Our network has reversible module and reconstruction net, so our loss includes outputting and reversible loss. The outputting loss is calculated as the L2 loss of $x_{out}$ - $x_{truth}$. The reversible loss calculation $x_{out}$ is mapped back to the CCD under the nature of the reversible optical path to obtain the L2 loss of the G ($x_{out}$) value to the actual measurement y. We defined the loss function as follows:

\begin{equation}
\setlength{\abovedisplayskip}{0pt} 
\mathcal{L} =\left \| \widehat{x}_{out}- x_{truth}  \right \|_{2}^{2}  + \xi \cdot \left \| \Phi\widehat{x}_{out}-x_{in} \right \|_{2}^{2}  
  \label{eq:17}
\end{equation}
where $\widehat{x} _{out}$ is the predicted values of the network, $\Phi $ represents the process of mask coding and dispersion of predicted values, $x_{in}$ is the measurement of CCD. $\xi$ is the penalty coefficient.

\begin{table*}[t]\scriptsize
\begin{center}
\caption{Comparisons between MLP-AMDC and SOTA methods on 10 simulation scenes (S1$\sim$S10) in KAIST. PSNR and SSIM are reported.}
\begin{tabular}{cccccccccccccccc}
\hline
& $\lambda$ & ADMM  & TSA & Gap  & DGS  & PnP-DIP  & BIR & HD  & MST & CST & Heros & DAUHST & SST & RDLUF & AMDC \\
& -net & -Net  & -net & -net  & MP  & -HSI  & NAT & Net  & ++ & -L & Net & -9stg & -LPlus & -9stage & -9stg \\

\hline
  PSNR & 31.77  & 33.58  & 32.30 & 24.36 & 32.63 & 31.26 & 37.58 & 34.97 & 35.34 & 36.12 & 34.45 & 38.36 & 39.16 & 39.57 & \pmb{48.61} \\
\hline
  SSIM & 89.0\%  & 91.8\% & 91.6\%  & 66.9\% & 91.7\% & 89.4\% & 96.0\% & 94.3\%  & 95.3\% & 95.7\% & 97.0\% & 96.7\% & 97.4\% & 97.4\% & \pmb{99.6\%}\\
\hline

\label{tab:1}
\end{tabular}
\end{center}
\end{table*}

\begin{figure*}[t]
  \centering
   \includegraphics[width=0.88\linewidth]{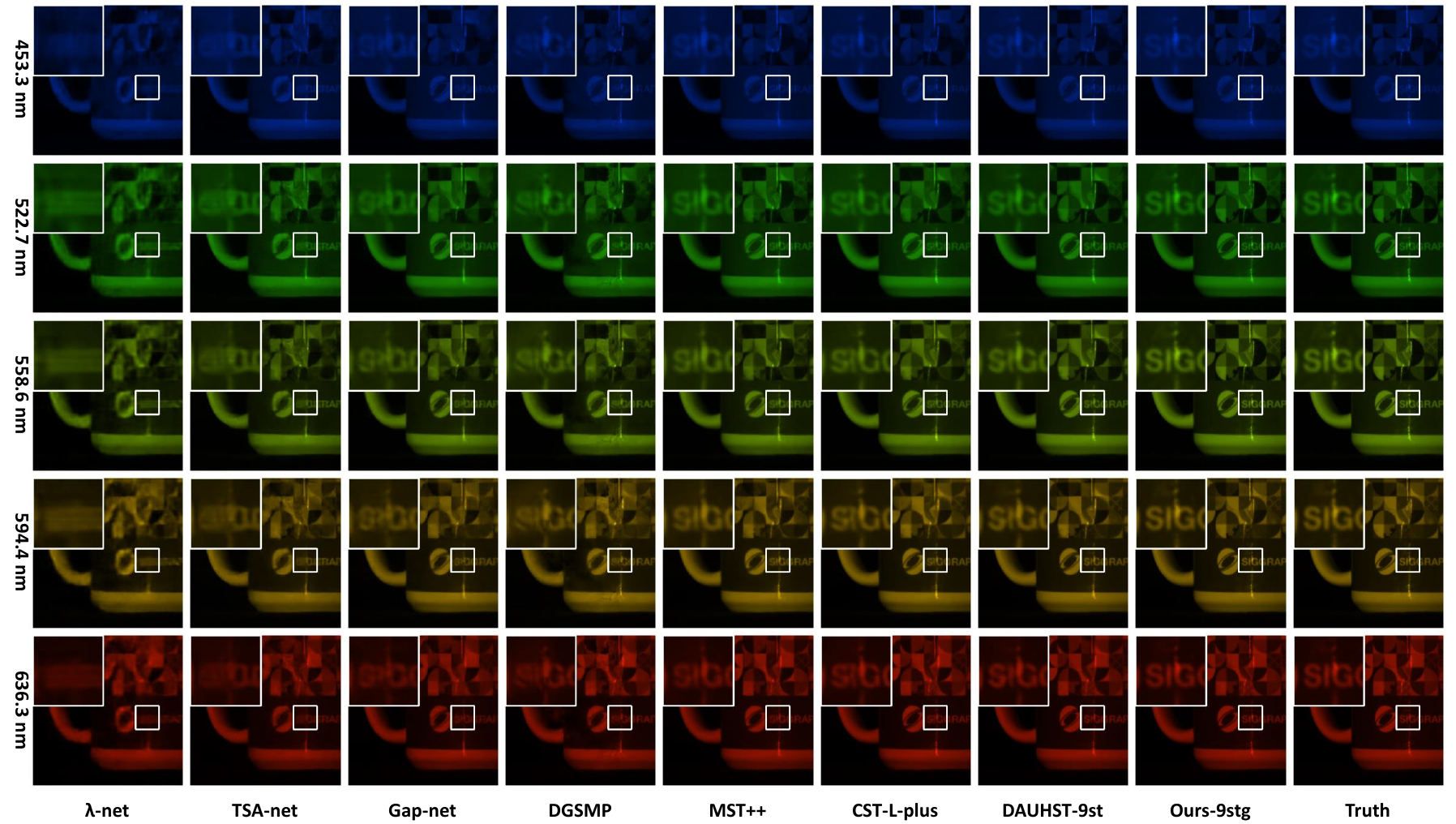}
   \caption{Visual comparisons of our MLP-ADAM and other SOTA methods of Scene 5 with 5 out of 28 spectral channels on the KAIST dataset.}
   \label{fig:6}
\end{figure*}

\begin{figure}[t]
  \centering
   \includegraphics[width=0.9\linewidth]{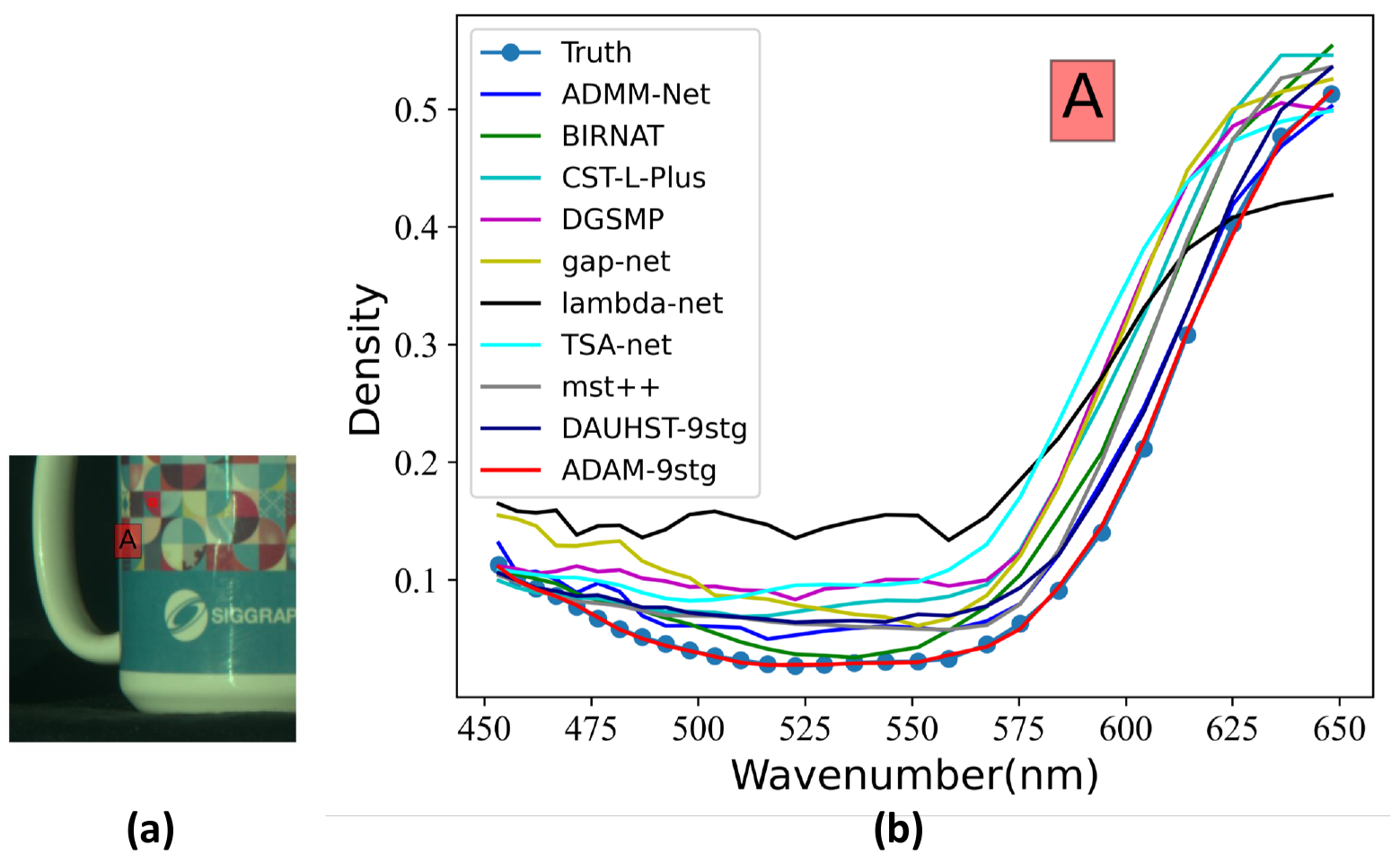}
   \caption{Spectral curves on the selected regions (A) and the visualization results show that our results have higher spectral accuracy and better perceptual quality.}
   \label{fig:5}
\end{figure}

\begin{figure}[t]
  \centering
   \includegraphics[width=1\linewidth]{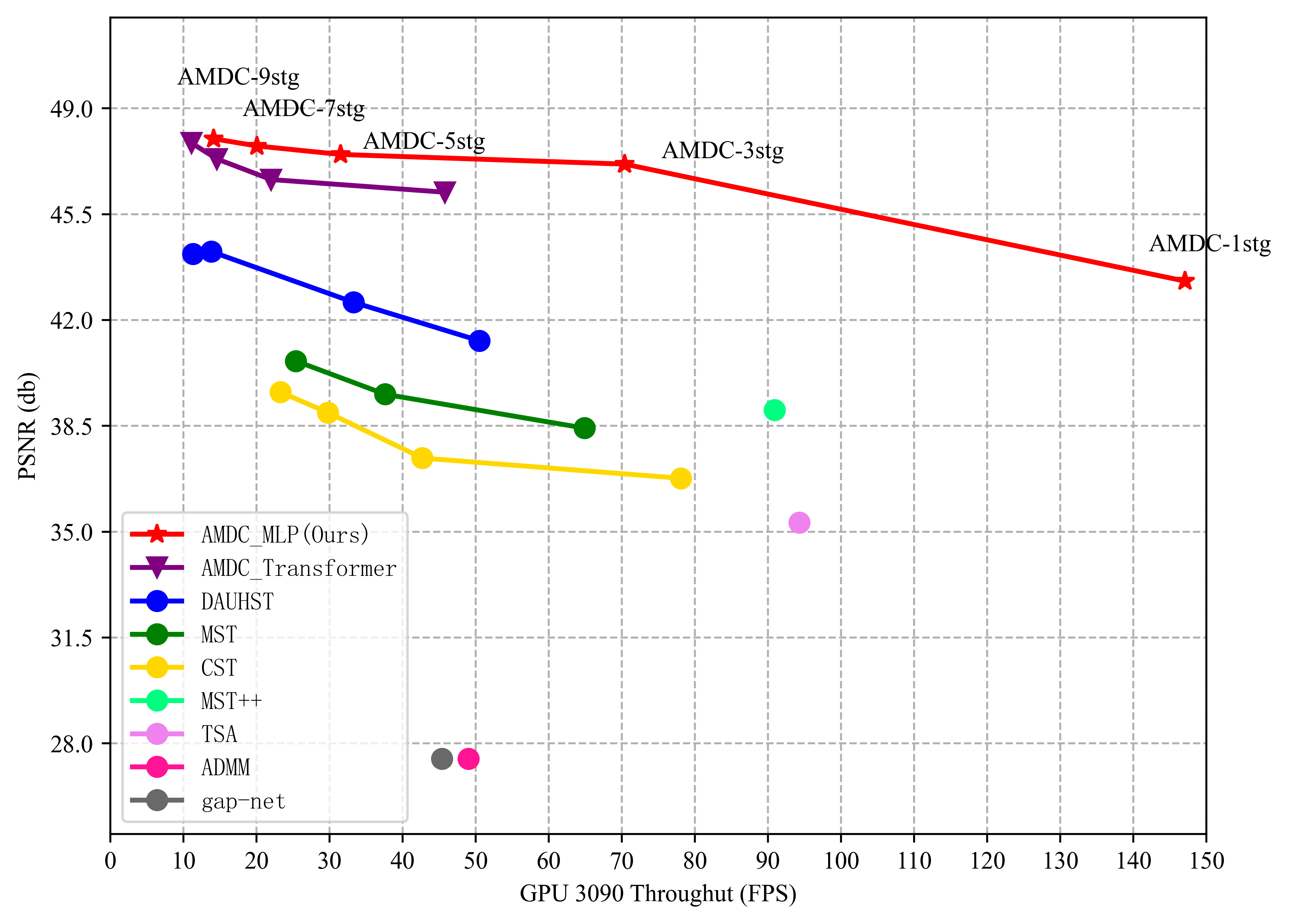}
   \caption{Comparison of FPS and PSNR for AMDCs and other SOTA methods.}
   \label{fig:8}
\end{figure}

\section{Experiments}

This section shows how the proposed SOTA reconstruction method outperforms others on numerous datasets, including re-evaluating the SOTA model in CASSI on the current dataset ARAD\_1K.

\subsection{Experiment Setup}

CAVE \cite{park2007multispectral} and KAIST's \cite{choi2017high} testing dataset has 28 spectral channels, ranging from 450 to 650 nm. CAVE and KAIST simulation hyperspectral image datasets are used. CAVE dataset has 32$\times$ 512 $\times$ 512 hyperspectral pictures. 30 hyperspectral photos from KAIST are 2704 $\times$ 3376. CAVE is the training set, following TSA-Net's schedule \cite{meng2020end}. Test 10 KAIST scenes.

ARAD\_1K presents a larger-than-ever natural hyperspectral image dataset. This data collection approximately doubles the ARAD\_1K HS data set to 1,000 pictures. ARAD\_1K has 31 hyperspectral images at 482 $\times$ 512 from 400 to 700 nm with a 10 nm increment. The 1,000 data set photos were divided: 900 training, 50 validation, and 50 test photos \cite{arad2022ntire}.

We select peak signal-to-noise ratio (PSNR), structural similarity (SSIM)\cite{wang2004image}, Mean Relative Absolute Error (MRAE) and Root Mean Square Error (RMSE) as the metrics to evaluate the HSI reconstruction performance. CAVE and KAIST continue to use PSNR, SSIM, parameters, and floating point numbers as evaluation measures. MRAE is ARAD\_1K's major metric.

We implement MLP-AMDC in Pytorch. All mrthods are trained on 1 $\times$ RTX 3090 GPU. Adam\cite{kingma2014adam} optimizer for 300 epochs ($\beta_{1} = 0.9$ and $\beta_{2} = 0.999$). The training rate starts at $4\times 10^{-4}$ and is halved every 50 epochs.

\begin{table*}[t]\footnotesize
\begin{center}
\caption{Comparisons between MLP-AMDC and SOTA methods on 50 scenes in ARAD\_1K. MRAE, RMSE, PSNR, SSIM , parameters, FLOPS and reconstruction time are reported.}
\begin{tabular}{ccccccccccccc}
\hline
method  & Avg MRAE & Avg RMSE     & Avg PSNR     & Avg SSIM     & Params(M)     & GFLOPS(G)     & Time for 50 pics (s)\\
\hline
$\lambda$-net & 0.1664 & 0.0228 & 33.29 & 91.63\% & 32.73 & 31.18 & 0.19  \\
\hline
TSA-net & 0.1392 & 0.0193 & 34.82 & 93.82\% & 44.30 & 142.34 & 0.53 \\

\hline
Gap-net & 0.3005 & 0.0466 & 27.02 & 84.26\% & 4.28 & 87.55 & 1.10 \\
\hline
Admm-net  & 0.2949 & 0.0475 & 27.03 & 84.22\% & 4.28 & 87.55 & 1.02 \\

\hline
MST-S  & 0.0984 & 0.0135 & 37.93 & 96.51\% & 1.12 & 14.93 & 0.77 \\

MST-M  & 0.0855 & 0.0119 & 39.02 & 97.21\%  & 1.81 & 20.85 & 1.33 \\

MST-L & 0.0776 & 0.0102 & 40.64 & 97.71\% & 2.45 & 32.18 & 1.97 \\

MST++  & 0.0936 & 0.0127 & 38.49 & 96.92\% & 1.63 & 23.88 & 0.55 \\

\hline
CST-S & 0.1308 & 0.0165 & 36.25 & 95.23\% & 1.46 & 12.21 & 0.64 \\

CST-M  & 0.1147 & 0.0153 & 36.91 & 95.77\% & 1.66 & 15.96 & 1.17 \\

CST-L  & 0.0956 & 0.0125 & 38.95 & 96.74\% & 3.66 & 27.17 & 1.68 \\

CST-LPlus  & 0.0854 & 0.0115 & 39.63 & 97.16\% & 3.66 & 33.95 & 2.15 \\

\hline
DAUHST-3stg & 0.0686 & 0.0094 & 41.31 & 97.90\% & 1.64 & 29.13 & 0.99 \\

DAUHST-5stg  & 0.0590 & 0.0076 & 43.89 & 98.36\% & 2.70 & 47.72 & 1.50 \\

DAUHST-7stg & 0.0575 & 0.0074 & 44.20 & 98.44\% & 3.77 & 66.32 & 3.62 \\

DAUHST-9stg & 0.0602 & 0.0073 & 44.18 & 98.45\% & 4.83 & 84.92 & 4.42 \\
\hline
AMDC-1stg  & 0.0710 & 0.0077 & 43.28 & 98.59\% & 1.25 & 25.81 & 0.34 \\

AMDC-3stg  & 0.0432 & 0.0046 & 47.15 & 99.34\% & 1.77 & 41.28 & 0.71 \\

AMDC-5stg & 0.0429 & 0.0046 & 47.47 & 99.34\% & 1.77 & 56.75 & 1.59 \\

AMDC-7stg & 0.0442 & 0.0045 & 47.75 & 99.32\% & 1.77 & 72.23 & 2.49 \\

AMDC-9stg & \pmb{0.0420} & \pmb{0.0044} & \pmb{47.99} & \pmb{99.35\%} & 1.77 & 87.80 & 3.54 \\

\hline
\label{tab:1}
\end{tabular}
\end{center}
\end{table*}

\subsection{results on CAVE and KAIST}
\textbf{(i)}On CAVE and KAIST, our best model AMDC-9stg yields very impressive results, i.e., 48.61 dB in PSNR and 99.6\% in SSIM ,which is more than 9 dB than the best PSNR of the SOTA published models, and the SSIM is more than 2.2\%. AMDC-9stg significantly outperforms RDLUF-9stage \cite{Dong_2023_CVPR}, SST-LPlus \cite{cai2023sst}, DAUHST-9stg \cite{cai2022degradation}, HerosNet \cite{zhang2022herosnet}, CST-L \cite{lin2022coarse}, MST++ \cite{cai2022mst++}, HDNet \cite{hu2022hdnet}, BIRNAT \cite{cheng2022recurrent}, PnP-DIP-HSI  \cite{meng2021self}, DGSMP \cite{huang2021deep}, GAp-net  \cite{meng2020gap}, TSA-net \cite{meng2020end}, ADMM-Net \cite{ma2019deep}, and $\lambda$-net \cite{miao2019net} of PSNR by 9.04, 9.45, 10.25, 14.16, 12.49, 13.27, 13.64, 11.03, 17.35, 15.98, 24.25, 16.31, 15.03 and 16.84 dB, and 2.2\%, 2.2\%, 2.9\%, 2.6\%, 3.9\%, 4.3\%, 5.3\%, 3.6\%, 10.2\%, 7.9\% , 32.7\% , 8.0\% , 7.8\% , and 10.6\% improvement of SSIM. Detailed comparison data are shown in Table 1, See appendix for detailed data.
\begin{table*}[t]\footnotesize
  \centering
  \caption{Evaluation of the effectiveness of different components.}
  \begin{tabular}{lccccccccc}
    \hline
    Base-line & Adaptive Mask & RGB & Noise Estimator & MRAE & RMSE & PSNR & SSIM & Params & GFLOPs  \\
     & & & & Avg & Avg & Avg(dB) & Avg & (M) &  \\
    \hline
    AMDC-3stg & $\surd$ & $\surd$ & $\times$ & 0.0459 & 0.0049 & 46.71 & 99.27\%  & 1.67 & 34.69 \\
    \hline
    AMDC-3stg & $\surd$ & $\times$ & $\surd$  & 0.0668 & 0.0070 & 43.65 & 98.71\% & 0.69 & 17.05 \\
    \hline
    AMDC-3stg & $\surd$ & $\surd$ & $\surd$  & 0.0432 & 0.0046 & 47.15 & 99.34\%  & 1.77 & 41.28 \\
    \hline
  \end{tabular}
  \label{tab:2}
\end{table*}

\begin{table*}[t]\footnotesize
  \centering
  \caption{Evaluation of the performance between MLP and transformer.}
  \begin{tabular}{lcccccccc}
    \hline
    Base-line & backbone & Avg MRAE & Avg RMSE & Avg PSNR & Avg SSIM & Params & GFLOPs & FPS \\
     & &  &  & (dB) &  & (M) &  & \\
     \hline
    AMDC-3stg & Transformer &  0.0488 & 0.0052 & 46.22 & 99.28\%  & 1.93 & 45.08 & 45.8 \\
    AMDC-5stg & Transformer & 0.0453 & 0.0049 & 46.64 & 99.33\% & 1.93 & 63.62 & 22.0 \\
    AMDC-7stg & Transformer & 0.0455 & 0.0048 & 47.32 & 99.31\%  & 1.93 & 82.15 & 14.6\\
    AMDC-9stg & Transformer & 0.0430 & 0.0044 & 47.85 & 99.38\%  & 1.93 & 100.69 & 11.1\\
    \hline
    AMDC-3stg & MLP &  0.0432 & 0.0046 & 47.15 & 99.34\%  & 1.77 & 41.28 & 70.4\\
    AMDC-5stg & MLP & 0.0429 & 0.0046 & 47.20 & 99.34\% & 1.77 & 56.75 & 31.5\\
    AMDC-7stg & MLP & 0.0442 & 0.0045 & 47.75 & 99.32\%  & 1.77 & 72.23 & 20.88\\
    AMDC-9stg & MLP & 0.0421 & 0.0044 & 47.99 & 99.35\%  & 1.77 & 87.80 & 14.12\\
    \hline
  \end{tabular}
  \label{tab:3}
\end{table*}

\begin{table*}[h]\footnotesize
  \centering
  \caption{Evaluation of the effectiveness of different types of Mask.}
  \begin{tabular}{lccccccccc}
    \hline
    Base-line & Mask type & Avg MRAE & Avg RMSE & Avg PSNR & Avg SSIM & Params & GFLOPs \\
     & &  &  & (dB) & & &  \\
     \hline
    AMDC-3stg & Manual Prior Mask &  0.0582 & 0.0067 & 44.03 & 98.81\% & 1.77 & 41.28 \\
    \hline
    AMDC-3stg & Rand Mask & 0.0504 & 0.0057 & 45.47 & 99.09\% & 1.77 & 41.28 \\
    \hline
    AMDC-3stg & Normal Mask & 0.0515 & 0.0056 & 45.60 & 99.13\% & 1.77 & 41.28 \\
    \hline
    AMDC-3stg & Aynamic Mask & 0.0432 & 0.0046 & 47.15 & 99.34\% & 1.77 & 41.28 \\
    \hline
  \end{tabular}
  \label{tab:3}
\end{table*}

\textbf{(ii)}It can be observed that our MLP-AMDC significantly surpass SOTA methods by a large margin while requiring much cheaper memory and computational costs. Compared with DU methods, AMDC-5stg outperforms RDLUF-9stage by 7.90 db but only costs 79.4\% (1.47/1.85) parameters and 39.6\% (47.59/120.08) FLOPs, outperforms DAUHST-9stg 9.11 db but only costs 23.9\% (1.47/6.15) parameters and 57.3\% (47.59/79.50) FLOPs. And outperforms other Transformer-based methods, our AMDC-1stg outperforms CST-L \cite{lin2022coarse} by 4.36 dB but only costs 34.6\% (1.04/3.00) Params and 79.7\% (22.18/27.81) FLOPs, and AMDC-1stg outperforms MST++ \cite{cai2022mask} by 4.87 dB, but only costs 28.4\% (1.04/3.66) Params and 78.8\% (22.18/28.15) FLOPs. In addition, our AMDC-3stg, AMDC-5stg, AMDC-7stg and AMDC-9stg outperforms other competitors by very large margins. we provide PSNR-parameters-FLOPs comparisons of different reconstruction algorithms in Fig. 1.

Fig. 5 plots the visual comparisons of our AMDC-9stg and other SOTA methods on Scene 3 with 5 (out of 28) spectral channels. The top-left part shows the zoomed-in patches of the white boxes in the entire HSIs, the reconstructed HSIs produced by AMDCs have more spatial details and clearer texture in different spectral channels than other SOTA methods. In addition, the spectral curves of the AMDCs have a higher correlation with the reference spectra in the quantitative spectral data. As shown in Fig. 6(b) with the red line and the blue line.

\subsection{Results on ARAD\_1K}
On ARAD\_1K, the comparison data of AMDC and SOTA models are shown in Table 2, our AMDC-9stg yields 0.0420 in MRAE, 0.0044 in RMSE, 47.99 in PSNR, 99.35\% in SSIM. AMDC-9stg still significantly outperforms DAUHST-9stg \cite{cai2022degradation}, CST-LPlus \cite{lin2022coarse}, MST++ \cite{cai2022mst++}, Admm-net \cite{ma2019deep}, Gap-net \cite{meng2020gap}, TSA-net \cite{meng2020end},  and $\lambda$-net \cite{miao2019net} of MRAE by 0.0182, 0.0434, 0.0516, 0.2529, 0.2585, 0.0972, and  0.1244, and of PSNR by  3.45, 8.36, 9.50, 20.96, 20.97, 13.17, and 14.70, and 0.90\%, 2.09\%, 2.43\%, 15.13\%, 15.09\%, 5.53\%, and 7.72\% improvement of SSIM, suggesting the effectiveness of our method.

\subsection{Ablation study}

\begin{figure}[t]
  \centering
  \includegraphics[width=1\linewidth]{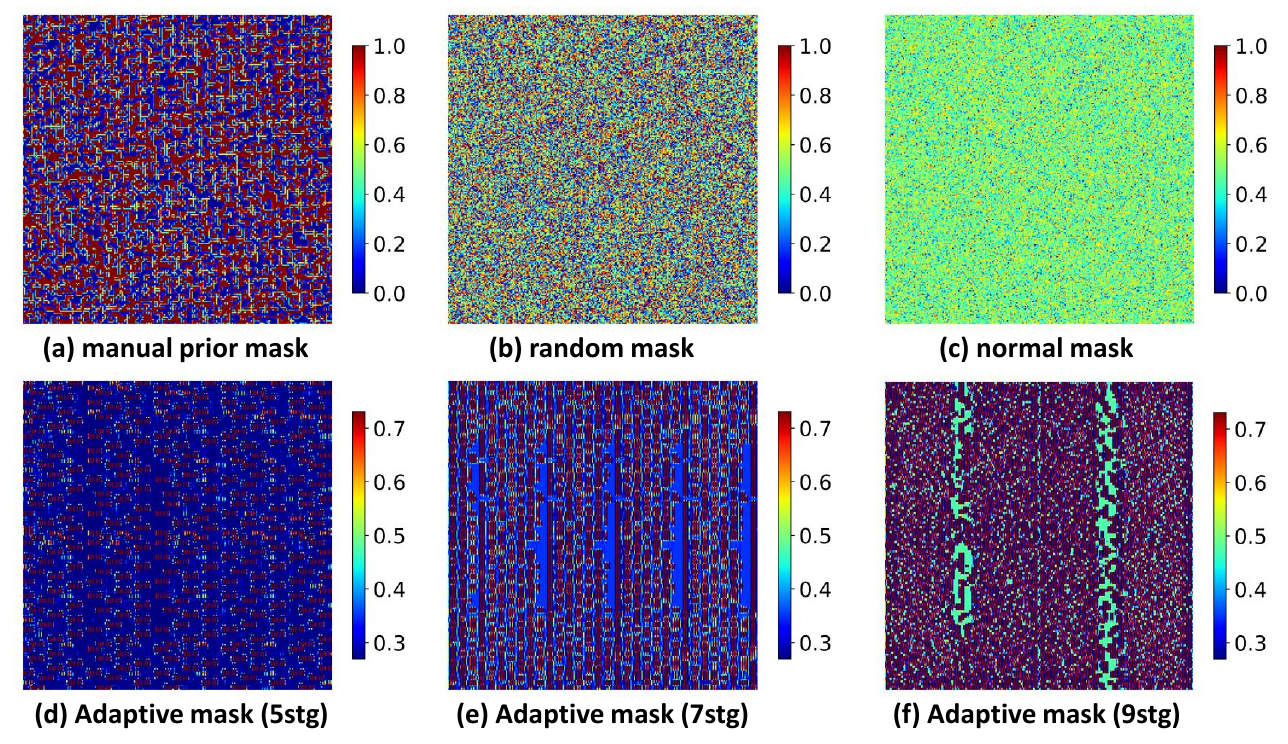}
  \caption{Comparison of manual prior mask, random mask, normal mask and dynamic mask.}
  \label{fig:7}
\end{figure}

Ablation analysis using ARAD\_1K datasets evaluates the proposed AMDCs' components. Adaptive mask, RGB branching, Noise Estimator, and MLP vs. transformer performance are our key focus. Tab. 3, tab. 4 and tab. 5 displays MRAE, RMSE, PSNR, and SSIM for RGB  ranching and Noise Estimator settings,  different AMDC sizes, and different mask methods.The adaptive mask and RGB branching increase the network equally. The difference between adaptive mask and other masks is shown in Fig. 8. Adaptive mask has a better encoding method based on RGB prior.

To evaluate the combined comparison of the reconstruction speed and performance of different algorithms on ARAD\_1K, we analyzed the PSNR and FPS of SOTA methods. As shown in Fig. 7, our AMDC-1stg is ten as fast for the same reconstruction quality, as well as other AMDCs.

\section{Conclusion}
This study proposes AMDC, a CASSI framework. The new approach addresses the problem of random and manual masks not fitting the dataset, which lowers reconstruction performance, and the transformer-based method is slower and more arithmetic. We present an MLP-AMDC with adaptive mask and multimodal reconstruction networks to integrate CASSI data with RGB photos using AMDC-CASSI. These unique methods provide very efficient MLP-AMDC models. Our technique outperforms SOTA algorithms even with cheaper settings and FLOPs.

\bibliography{aaai24}

\end{document}